\documentclass[twoside, 11pt]{article} 

\input epsf

\def\IRINI#1#2#3#4#5{{%
\hbox to 0pt{\vbox to 0pt{\hbox to \textwidth{\vbox to \textheight{%
\noindent\parbox{\textwidth}{
\setlength{\epsfxsize}{\textwidth}
\centerline{\epsfbox{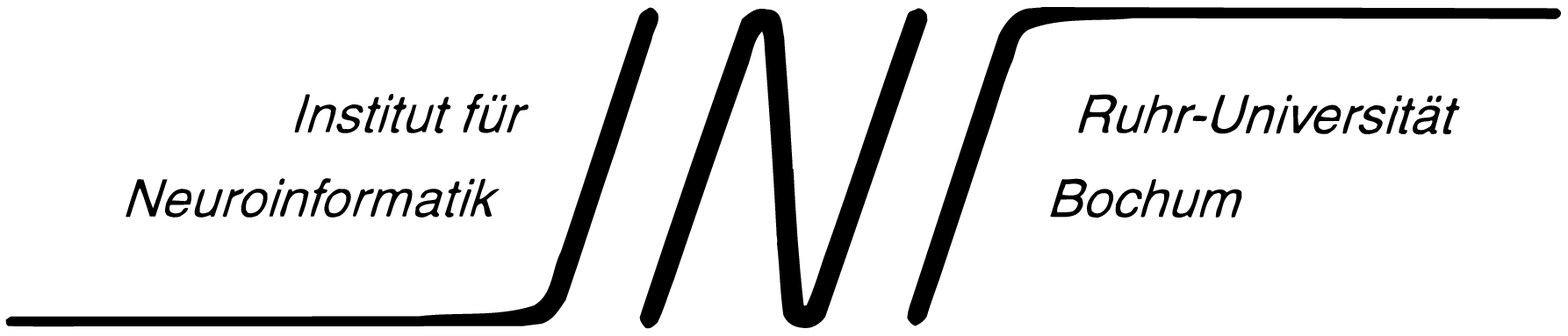}}
}\\[3mm]
\noindent\parbox{\textwidth}%
{\begin{center}
\Large Internal Report #3
\end{center}}
\vspace*{4cm}
\noindent\parbox{\textwidth}%
{\begin{center}
\Large\bf #1
\end{center}}
\vspace*{1cm}
\noindent\parbox{\textwidth}%
{\begin{center}
{\large\em by}\\[2mm]
\large #2
\end{center}}
\vfill
\noindent%
\centerline{\parbox{6cm}%
{Ruhr-Universit\"at Bochum\\
Institut f\"ur Neuroinformatik\\
44780 Bochum}%
\hfill%
\parbox{3cm}{\rule[-1cm]{0cm}{2.5cm}\parbox{4cm}%
{\centering
\setlength{\epsfxsize}{2.8cm}
\epsfbox{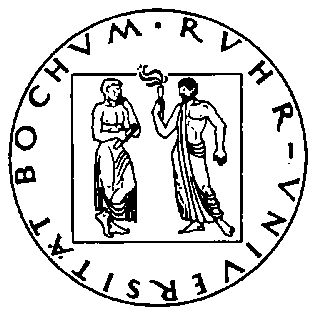}
}}%
\hfill%
\parbox{6cm}%
{\begin{flushright}
IR-INI #3\\
#4 \ #5\\
ISSN 0943-2752
\end{flushright}}}
\begin{center}
\copyright \ #5 Institut f{\"u}r Neuroinformatik, %
Ruhr-Universit{\"a}t Bochum, FRG
\end{center}}}\vss}\hss}}%
\setcounter{page}{0}
\pagestyle{myheadings}
\markboth{IR-INI #3, \copyright \ #5 Institut f{\"u}r Neuroinformatik, %
Ruhr-Universit{\"a}t Bochum, FRG}{IR-INI #3, \copyright \ %
#5 Institut f{\"u}r Neuroinformatik, Ruhr-Universit{\"a}t Bochum, FRG}
\markright{IR-INI #3, \copyright \ #5 Institut f{\"u}r Neuroinformatik, %
Ruhr-Universit{\"a}t Bochum, FRG}
\thispagestyle{empty}
\newpage}

\usepackage{graphicx}
\usepackage{epsfig}
\usepackage{wrapfig}
\usepackage{boxit}

\newtheorem{definition}{Definition}

\begin{document}

\IRINI%
{Evolution in time-dependent fitness landscapes}
{Claus~O.~Wilke}
{98--09}
{November}
{1998}

\title{\bf Evolution in time-dependent fitness landscapes}
\author{Claus~O.~Wilke\thanks{email claus.wilke@ruhr-uni-bochum.de}}
\date{}
\maketitle
\footskip 1.5cm

\begin{abstract}
Evolution in changing environments is an important, but 
little studied aspect of the theory of evolution. The idea of adaptive
walks in fitness landscapes has triggered a vast amount of research
and has led to many important insights about the progress of
evolution. Nevertheless, the small step to time-dependent fitness
landscapes has most of the time not been taken. In this work,
some elements of a theory of adaptive walks on changing fitness
landscapes are proposed, and are subsequently applied to and tested on
a simple family of time-dependent fitness landscapes, the oscillating
$NK$ landscapes, also introduced here. For these landscapes,
the parameter governing the evolutionary dynamics is the fraction of
static fitness contributions $f_{\rm S}$. For small $f_{\rm S}$, local
optima are virtually   non-existent, and the adaptive walk constantly
encounters new genotypes, whereas for large $f_{\rm S}$, the
evolutionary dynamics reduces to the one on static fitness
landscapes. Evidence is presented that the transition between the two
regimes is a 2nd order phase transition akin a percolation
transition. For $f_{\rm S}$ close to the critical point, a rich
dynamics can be observed. The adaptive walk gets trapped in noisy
limit cycles, and transitions from one noisy limit cycle to
another occur sporadically.
\end{abstract}

\section{Introduction}
The key element of Darwinian evolution is natural selection, which
prefers advantageous mutations and rejects deleterious ones.
It is possible to separate any evolutionary system under study into
two components: a single species (or population of
individuals) and an environment. The individuals in the population
change because of mutations, and the environment discriminates between
better adapted ones (the ones generating more offspring, i.e., the
``fitter'' ones) and  worse adapted ones. The difference in 
the number of offspring suffices to remove efficiently the worse
adapted individuals 
from the population. Clearly, the environment does not
necessarily remain constant in time. It consists of all the influences
acting on the species, and it therefore contains a-biotic and biotic
components. Changes may have an
a-biotic origin (e.g., change in mean temperature, change in sea level) or
a biotic origin (e.g., other species in the same ecosystem changing
because of adaption). The latter case is usually referred to
as coevolution. From the viewpoint of a single species that adapts in
an environment, all evolutionary scenarios can be divided into three
distinct classes.

\begin{wrapfigure}[28]{r}{.55\columnwidth} 
\begin{tripbox}
\includegraphics[width=.5\columnwidth]{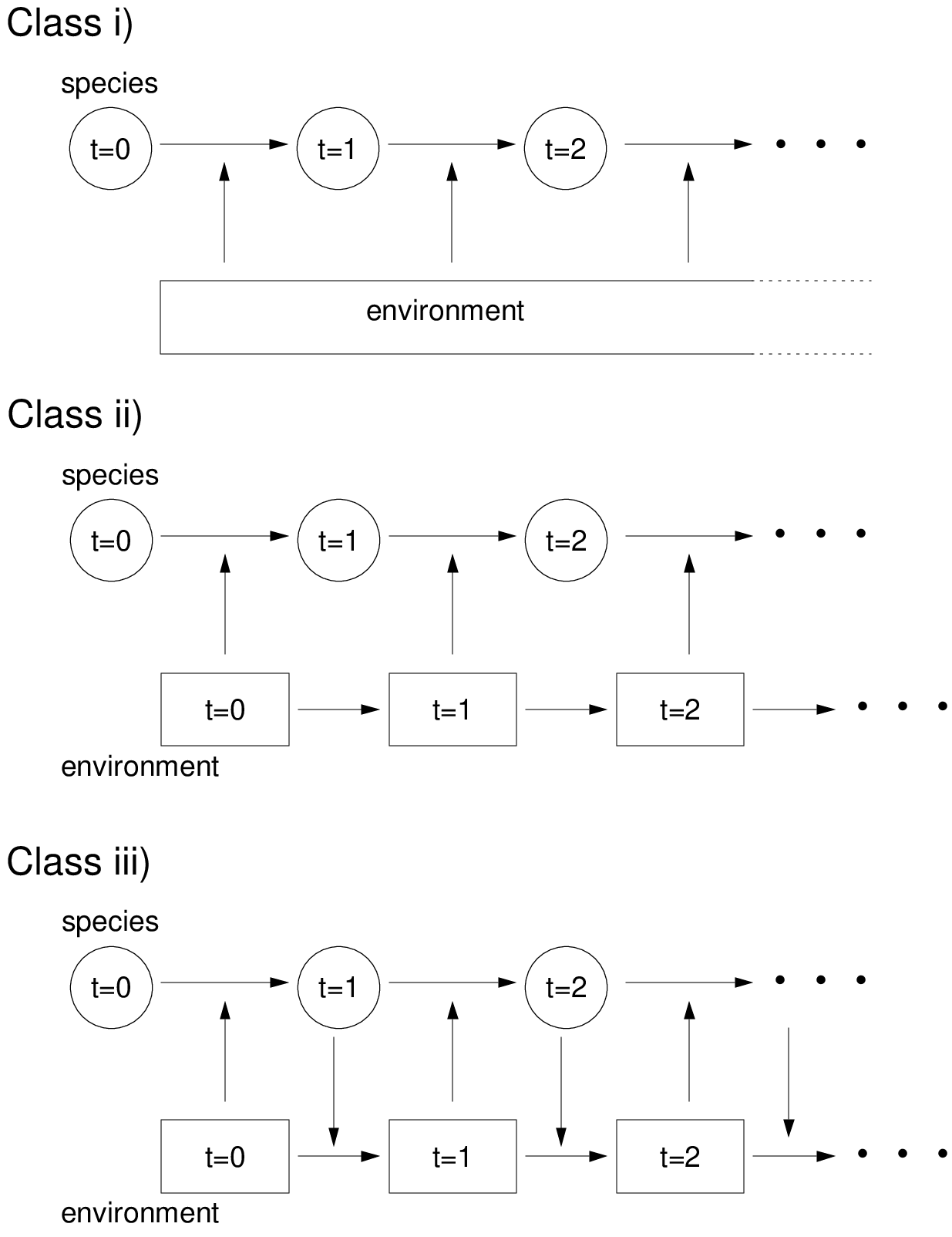}
\caption{Evolutionary scenarios can be divided into three classes: i)
  evolution in a constant environment, ii) evolution in a variable
  environment without feedback, and iii) evolution with feedback to
  the environment.}
\label{fig:evol_classes}
\end{tripbox}
\end{wrapfigure}

These classes are, as depicted in
Fig.~\ref{fig:evol_classes}, 
i) evolution in a constant environment, ii) evolution in a variable
environment without feedback to the environment, and iii) evolution
with feedback to the environment. The last class is the coevolutionary case
mentioned above. The current knowledge about evolutionary dynamics is
very unevenly distributed over these  three classes. An overwhelming
amount of theoretical results is known 
for class i). Most of the work
in the field of population genetics has been done with constant
environment (e.g.~\cite{Ewens79,Gale90,Kimura84}), and also the idea of
adaptive walks in fitness landscapes, introduced by
Wright~\cite{Wright32}, and intensively studied for example by
Kauffman~\cite{Kauffman92}, has usually only
been applied to constant
fitness landscapes. Class iii) has also been studied
intensively. One the one hand, there exists a rich theoretical
ecology literature (e.g.~\cite{Freedman80,HallamLevin86}), dealing
with predator-prey interactions such as
Lotka-Volterra systems, or with host-parasite interactions. On the
other hand, over the last decade, coevolutionary scenarios have been studied 
through direct competition of two or more species in Artificial Life style
computer simulations (e.g.~\cite{Hillis91,KauffmanJohnsen91,Reynolds94}). The
evolutionary patterns observed there are often very complex, and a
clear theoretical understanding of these patterns has not been 
obtained yet. Finally, Class ii), which lies between classes
i) and iii), and which could therefore yield many new insights about
the connection between adaptive walks and coevolution, has
nevertheless widely been neglected. There exists some work  on
evolution in randomly fluctuating environments\cite{Gillespie91}, but
this work  obviously cannot cover the wide range of possible 
scenarios within class ii). In particular, a theory for
the natural generalization of class i) adaptive walks, i.e., adaptive
walks on a changing fitness landscape, is still missing.

\section{Environmentally guided drift}
\label{sec:env_guided_drift}

Evolutionary processes in static fitness landscapes tend to get stuck quickly
in local optima, often far away from some globally optimal
solution. This hinders evolution's progress, and
severely restricts the amount of possible solutions that can be found
during the adaptive progress. Nevertheless, Nature's rich diversity of
extremely clever solutions to a large array of different problems
indicates that some mechanisms are driving evolution
out of local optima. Probably, the most common proposal for such a
mechanism is the idea of neutral evolution, introduced by
Kimura~\cite{Kimura84}, and based on the observation that the
genotype-phenotype mapping of 
living beings discards information.
Certain mutations can appear without change in the organism's fitness, and 
therefore, without being constrained by natural selection. The accumulation of
such neutral mutations is called neutral drift. In principle, the
neutral drift could allow a certain gene to drift far away from its
original position in genome space, and then to change to a very much
advantageous state by means of a single point mutation. There exists a
lot of interesting theoretical results on this
topic~\cite{FontanaSchuster98,HuynenStadlerFontana96,NewmanEngelhardt98,ReidysStadlerSchuster97}. However, on the 
level of proteins, there is not much evidence in favor of this
viewpoint~\cite{BennerEllington88,KreitmanAkashi95}. Clearly, there are neutral
mutations on the level of DNA
(different codons code for the same amino acid) or in the RNA folding
problem, but on the level of amino acid sequences almost every change
seems to have some selective impact. 

One out of the numerous investigations demonstrating the absence of neutrality
on the protein level is the following study of the
\emph{Fast/Slow} 
(\emph{Adh-F/Adh-S}) polymorphism at the alcohol dehydrogenase locus
(\emph{Adh}) of \emph{Drosophila melanogaster}. The \emph{Fast/Slow}
polymorphism is probably the most intensively studied polymorphism at
all, with a number of 364 scientific research papers written on it
already back in 1988~\cite{Chambers88}.
Kreitman~\cite{Kreitman83} sequenced eleven cloned \emph{Drosophila m.} 
\emph{Adh} genes from five natural populations. Forty-two out of 43
differences in the DNA sequence were silent, only one base-pair
substitution actually caused a change in the amino acid sequence. This
base-pair substitution constituted exactly the \emph{Adh-F/Adh-S}
polymorphism. The
immediate conclusion that could be drawn was that most amino acid changes in
alcohol dehydrogenase would be selectively deleterious for the
wild-type \emph{Drosophila m}. Other studies demonstrate that the
\emph{Fast/Slow}  polymorphism in itself is sensitive to some selective
pressure. The population's composition of \emph{Adh-F} and
\emph{Adh-S} mutants was found to change with the geographical
latitude (latitudinal cline)~\cite{Sampsell77,Wilksetal80}. In a
recent study, Berry and Kreitman~\cite{BerryKreitman93} found a cline
along the East Coast of North America that could only be observed in
the amino acid responsible for the \emph{Adh-F/Adh-S} polymorphism,
but not in the accompanying silent mutations. This finding rejects
neutral drift as the origin of the cline.

Comprehensive reviews on the subject of protein evolution's
sensitivity to selective pressures can be found
in~\cite{BennerEllington88,KreitmanAkashi95}. 

Another mechanism capable of driving evolution away from local optima is
a slowly changing environment, as proposed by Benner and
Ellington~\cite{BennerEllington88}. If most changes in the amino acid
sequence of 
a certain protein are
not selectively neutral, then, on the other side, most changes in the
selective force will have some impact on the amino acid sequence of
this protein. A slowly changing environment will then (on an
evolutionary time scale) drag genes around in genotype space. This
results in another form of genetic drift, which can happen under
complete absence of neutrality. However,
 this drift is not a diffusion process such as
neutral drift. Instead, if the environmental changes happen slowly, the
population as a whole will move through the genotype space, since
transitions to a selectively advantageous state happen very fast, as
first order phase transitions~\cite{Adami98}. Such environmental
influences on the evolutionary dynamics have rarely been considered,
and the kind of dynamics they will produce is not at all
understood. In the reminder of this paper, we will investigate the
influence of a slowly changing environment for a simple example of
a time-dependent fitness landscape. Note that we will only be
interested in environmental changes happening on time-scales much
larger than the population's typical transition times to states with higher
fitness. Therefore, we can restrict our investigations to the study of
adaptive walks~\cite{Kauffman92}.  But, before we embark upon this
endeavor, we have to define some useful notions.

First of all, we are going to define the concept of environmentally
guided drift of a 
gene. This concept will be useful in the further study of evolution in
time-dependent fitness landscapes.
\begin{definition}[Environmentally guided drift] A gene is said to
  drift environmentally guided if in every jump from one point in
  genotype space to another, the gene's fitness in this time step
  remains either constant or increases.
\label{def:env_guided_drift}
\end{definition}
This definition includes neutral drift as a special case, and it also
includes those adaptive walks in which transitions to lower fitnesses are
forbidden. Note that the fitness can nevertheless decrease
during environmentally guided drift because of environmental
changes. However, this decrease is completely due to a change in the
structure of the underlying fitness landscape, and not to a
movement in the fitness landscape. 

Neutral drift has been included in the 
Definition~\ref{def:env_guided_drift} to make the term environmentally
guided drift sufficiently general. However, our main interest in this
paper lies in fitness landscapes that do not contain any neutrality.

Definition~\ref{def:env_guided_drift} reads very much like the
definition of standard hill-climbing, and is of course nothing else in
essence. Nevertheless, it seems appropriate to introduce a new notion
to remind ourselves that the dynamics changes drastically in a
time-dependent fitness landscape. In particular, a gene may appear to
be in a local optimum for almost every single time step, and
nonetheless it may travel constantly due to the changing positions of
the local optima. 

As in the case of pure neutral drift, the environmentally guided drift
could allow a gene to arrive at a point in 
genotype space that would not have been accessible by means of simple
hill-climbing on a fixed fitness landscape. Since environmental changes
are often oscillatory (seasons, for example, happen on an evolutionary
time scale 
for microorganisms) or pseudo-oscillatory (for example, cf. Fig. 2
in~\cite{Rietti-ShatiShemeshKarlen98}), a species most likely
encounters the same environmental state several times, and can use the
intermediate times of environmental guided drift for further adaption.

\section{Environmentally linked networks}

An important ingredient of the neutral theory of evolution is the
notion of neutral networks. A neutral network is formed by points in
the genotype space that have identical fitnesses and that are connected
by elementary mutations. The neutral drift is a diffusion process over
these networks. In analogy to the situation of neutral evolution, we can
define structures on which the environmentally guided drift naturally
takes place. These structures will in the following be called
environmentally linked networks or short EL networks.


EL networks can be defined on any time-dependent fitness landscape
that has  identical states, i.e., on fitness landscapes
for which there exists a sequence of times $t_0, t_1 ,
t_2, \dots$\,, such that 
\begin{equation}\label{eq:FL_equal_states}
  {\cal F}(t_0) =  {\cal F}(t_1) =  {\cal F}(t_2) = \dots\,, 
\end{equation}
where  ${\cal F}(t) =  {\cal F}(t')$  means that all points in genotype
space have the same fitness at times $t$ and $t'$. 

\begin{definition}[Environmentally linked networks]
\label{def:EL_networks}
An environmentally linked network is the set of all those points in
the genotype 
space that can be reached by means of environmentally
guided drift at times $t_i$, for which ${\cal F}(t_0) =  {\cal
  F}(t_i)$, starting from some fixed genome at time $t_0$.
\end{definition}


The EL networks according to Definition~\ref{def:EL_networks} contain
all the points in the genotype space that can in principle be reached
by sole hill-climbing. The probability to reach some of these points
may be very low. Therefore, these points will most certainly never be
encountered. As a consequence, EL networks are too ``large'' for
practical purposes, they contain many points with no relevance for the
evolutionary dynamics. To resolve this problem, we define smaller
structures, in which the most unlikely points are removed.

\begin{definition}[$\epsilon$-environmentally linked networks]
\label{def:epsilon-EL_networks}
The set of all the points $x_i$ in an EL network 
for which  a sequence of points
$x_0, x_1, \dots, x_{i-1}, x_i$ in
the EL network exists, such that every transition from $x_{j-1}$ to
$x_j$ can occur with probability 
$\epsilon^{t_j-t_{j-1}}$, is called an $\epsilon$-environmentally
linked network. 
\end{definition}
This definition immediately gives a trivial upper bound to the number of
points $n_{t}$ that can belong to the $\epsilon$-EL network after 
$t$ evolution time steps:
\begin{equation}
  n_t \leq \frac{1}{\epsilon^t}\,.
\end{equation}

A
question that  arises in conjunction with the above definitions
is whether environmentally guided drift 
can produce, in absence of neutrality, analogues of percolating
neutral networks, i.e., EL networks 
whose points are close in Hamming-distance to any arbitrary point in
the genotype space. The existence, or non-existence, of percolating
networks determines whether a mechanism is efficient in generating
evolutionary innovation.

\section{Oscillating $NK$-models}
\label{sec:oscillatingNK}

Environmental changes can be modelled in many different ways. For example,
the fluctuations can be stochastic or deterministic, continuous or
discontinuous, spatially resolved or unresolved. Clearly, each of the
different possibilities has its reason of being. Gillespie has studied
the effects of selection in a randomly fluctuating
environment~\cite{Gillespie91}. In his models, the fluctuations'
time scale is comparable to the duration of one generation of the organisms
considered. As a result, the genotype with the largest geometric mean
fitness dominates the population, and ultimately wins over all
competing genotypes. In the present work, we are interested in a
different situation. The subject of this work are environmental
changes that happen so slowly that the population as a whole moves
to a more advantageous genotype once it is discovered. In this
situation, a randomly fluctuating environment does not seem to be
appropriate. Rather, we are going  to consider a
continuous, deterministic model. For reasons of simplicity, we 
will also neglect any spatial structure. A changing environment of
this kind can easily be included into the
well-known $NK$ family of fitness
landscapes~\cite{Kauffman92,KauffmanWeinberger89}, standard
landscapes commonly used for the  study of  adaptive walks.
An $NK$-landscape is a 
fitness landscape for a bitstring of $N$ bits. The fitness of a
bitstring is the average over the fitness contributions of all bits in
the string. The contribution of a single bit depends on its state and
on the state of $K$ other bits interacting with it.  Mathematically, we
can write the fitness $f$ as:
\begin{equation}
  f = \frac{1}{N} \sum_{i=1}^{N} f_i(S_i, S_{\alpha_{i1}}, \cdots, 
                        S_{\alpha_{iK}})\,,
\end{equation}
where $S_i$ is the state of the $i$th bit (zero or one), $\alpha_{ij}$
is the number of the $j$th bit interacting with bit $i$, and
$f_i(\cdot)$ is a function that maps the state of the $i$th bit and
its interacting bits onto a real number between~0 and~1. In the
following, we will write the state of the $K+1$ bits $S_i,
S_{\alpha_{i1}}, \cdots, S_{\alpha_{iK}}$ as $\{S\}_i$. Usually, the
functions $f_i(\{S\}_i)$ are realized as tables containing a different
fitness contribution $C_\alpha$ for every possible arrangement of bits:
\begin{equation}\label{eq:fitness_contrib}
  f_i(\{S\}_i)= C_{i,\{S\}_i}\,.
\end{equation}
A straightforward generalization of this concept is reached by making
the functions 
$f_i(\{S\}_i)$ time dependent. Instead of the constants on the
right-hand side of Eq.~(\ref{eq:fitness_contrib}) we then have to use
time-dependent functions:
\begin{equation}
 f_i(\{S\}_i,t) = F_{i,\{S\}_i}(t)\,.
\end{equation}
The functions $ F_{i,\{S\}_i}(t)$ could in principle be arbitrarily
complicated. Here, we are going to consider a simple trigonometric time
dependency, i.e., 
\begin{equation}
F_{i,\{S\}_i}(t)=\frac{1}{2}[\sin(\omega_{i,\{S\}_i} t +
\delta_{i,\{S\}_i})+1]\,.
\end{equation}
This introduces only a single additional 
constant per fitness contribution, since there are now two of them
(the frequency $\omega_{i,\{S\}_i}$, and the phase
$\delta_{i,\{S\}_i}$). The frequencies $\omega_\alpha$ and the phases
$\delta_\alpha$ are chosen randomly at the beginning of the
evolutionary process, and then kept fixed for all times $t$. The
phases have to be distributed uniformly in the interval 
$[0; 2\pi)$ if the resulting fitness landscape is supposed to be
homogeneous in time. The choice of the frequencies determines whether
the fitness landscape is periodic in time. If we set the frequencies
to $\omega_{i,\{S\}_i}=2\pi n_{i,\{S\}_i}/T$, with
$n_{i,\{S\}_i}$ being integral, and $T$ being arbitrary, but the same for all
$\omega_{i,\{S\}_i}$), we 
obtain a periodic fitness landscape with oscillation period
$T$. Throughout the rest of this paper, we will stick to this choice. Its
advantage rests in the easy comparison of a bistring's evolution in different
oscillation periods.

If we set a frequency $\omega_{i,\{S\}_i}$ to zero, the according fitness
contribution $F_{i,\{S\}_i}(t)$ is constant in time. 
The fraction of these non-oscillatory (static) fitness contributions
will be called $f_{\rm S}$ in the following. The quantity $f_{\rm S}$
is the total number of static fitness contributions, divided by the total
number of all fitness contributions. Adaptive
walks on  oscillating $NK$-landscapes show several distinct classes of
behavior, most strongly influenced by $f_{\rm S}$. 

\subsection{Numerical experiments}

\begin{figure}
\begin{center}
\includegraphics{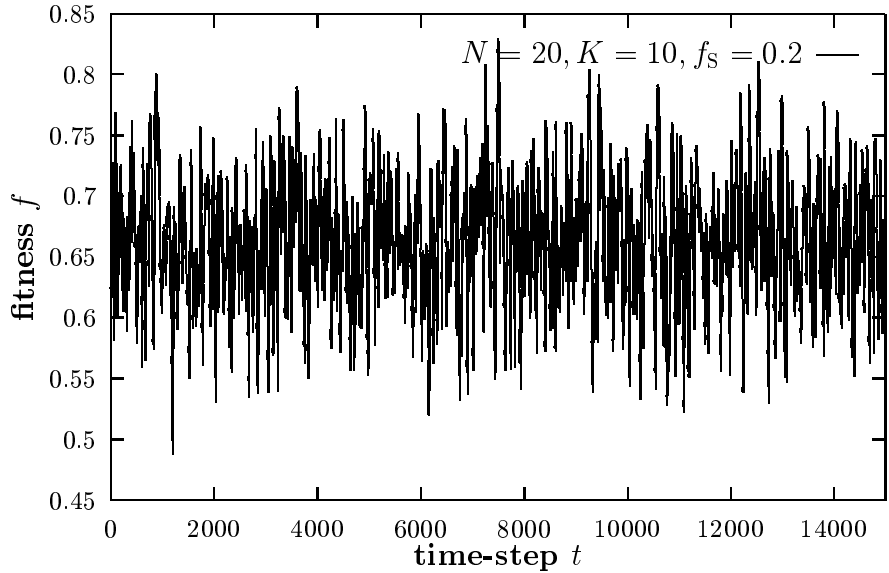}
\caption{The evolutionary dynamics is chaotic for small $f_{\rm S}$.}
\label{fig:OscNK1}
\end{center}
\end{figure}

\begin{figure}
\begin{center}
\includegraphics{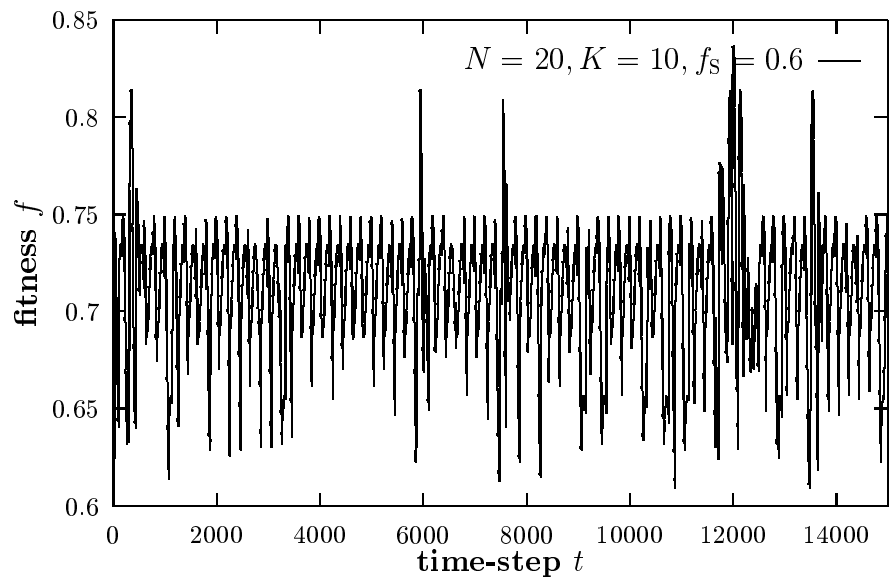}
\caption{With increasing $f_{\rm S}$, some bits in the bitstring
  freeze out, and the evolutionary pattern becomes more and more oscillatory.}
\label{fig:OscNK2}
\end{center}
\end{figure}

\begin{figure}
\begin{center}
\includegraphics{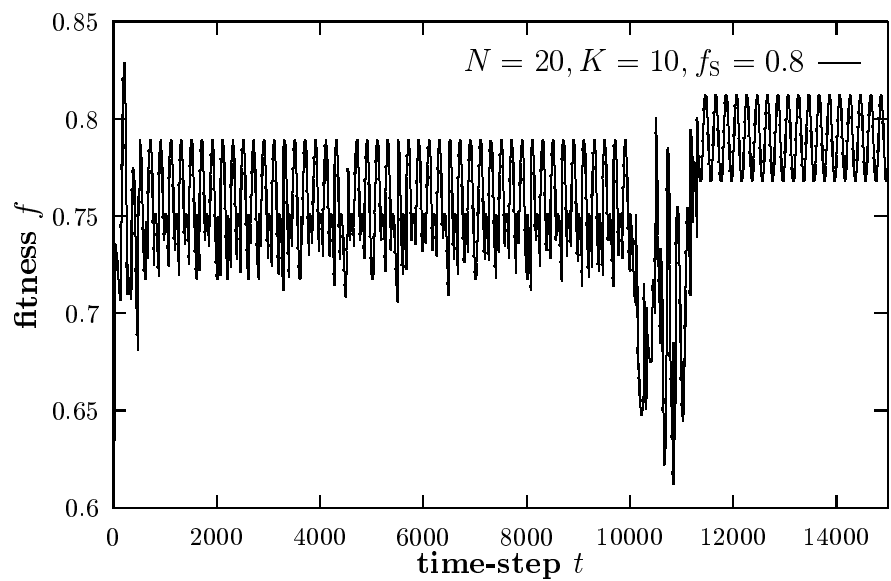}
\caption{A transition from one oscillatory state to
  another, with higher mean fitness.}
\label{fig:OscNK3}
\end{center}
\end{figure}

We have done several numerical experiments with oscillating $NK$-landscapes.
In all simulations presented below, only one 
oscillating mode was used. This means, all frequencies
$\omega_i$ were set either to zero or to some fixed value $\omega$. If
not indicated 
otherwise, the value $\omega=2\pi/200$ was used. In Figures \ref{fig:OscNK1}-\ref{fig:OscNK3}  some typical
runs of adaptive walks in oscillating $NK$-landscapes are presented. The
adaptive walk 
was performed exactly as in Kauffman's original work: a random point mutation
was accepted if it increased the bitstring's fitness. Otherwise, the
mutation was rejected. Hence, the adaptive walk per definition took
place on an EL network. 
Fig.~\ref{fig:OscNK1} shows an example of the 
evolutionary dynamics with a
relatively low fraction of static fitness contributions. The resulting
pattern is a chaotically changing fitness. With almost every accepted mutation,
a new genotype is encountered. The environmentally guided drift resembles 
a random walk on the fitness landscape. Interestingly, this is the
case even for small $K$. $NK$-landscapes with small $K$ are highly
correlated, which should bias the adaptive walk. Nevertheless,
it seems the oscillations destroy the correlations' effect. The
picture changes drastically with increasing $f_{\rm S}$. The higher
amount of static fitness contributions reduces the number of
possible advantageous mutations in every time step. The sites connected
to static contributions freeze out in a locally optimal state, and
only the sites connected to oscillating contributions can still
change. Hence, the dynamics gets confined in a small region of the
genotype space. The same mutational patterns are seen over  and over again
in the different oscillation periods. In the fitness plots, we can
identify this behavior with a periodic or almost periodic change of
the fitness, as shown in Fig.~\ref{fig:OscNK2}. Using the language of
dynamic systems, we can say that the attractor of an adaptive walk
on an oscillating landscape with intermediate $f_{\rm S}$ is a limit
cycle. Since the dynamics we are investigating here is not
deterministic, the limit cycles we observe will generally be
noisy. Hence, the 
limit cycles have to be understood in the sense of the $\epsilon$-EL
networks defined 
above. With a certain small probability $p<\epsilon$, the process can
leave an attractor, and find a new one. A
transition from one attractor to another is shown in
Fig.~\ref{fig:OscNK3}.  
 In this case, the mean fitness has significantly
increased after the transition.

A detailed analysis of the transition in Fig.~\ref{fig:OscNK3}
reveals the typical behavior of noisy limit cycles in oscillating
$NK$-models. We begin our analysis at time 
$t=6000$. The dynamics proceeded as follows. Underlined bits have
been mutated in the corresponding time step.
\begin{center}

\ttfamily
\begin{tabular}{ccc}
 \textrm{Time $t$} & \textrm{Bitstring}  & \textrm {Fitness $F(t)$}\\
 6000 & 10001101000000010100  & 0.7495\\
 6001 & 10001101000000010100 & 0.7497  \\
 6002 & 10001101000000010100  & 0.7502 \\
\vdots & \vdots & \vdots \\
6051 & 10001101000000010100 & 0.7364 \\
6052 & 1000\underline{0}101000000010100 & 0.7396 \\
\vdots & \vdots & \vdots \\
6185 & 1000\underline{1}101000000010100 & 0.7433 \\
\vdots & \vdots & \vdots \\
6262 & 1000\underline{0}101000000010100 & 0.7505 \\
\vdots & \vdots  & \vdots\\
6379 & 1000\underline{1}101000000010100 & 0.7399 \\
\vdots & \vdots & \vdots \\
\end{tabular}
\end{center}
We find a limit cycle with two states. At the beginning of each
period, bit 5 is switched from \texttt{0} to \texttt{1}. Towards
the end of the period, the bit is switched back to \texttt{0}. Remember we
used $T=200$ in this run, hence oscillation periods started at $t=6000,
6200, 6400, \dots$. 
The exact times in which bit 5 was switched differ from period to
period, because of the nondeterministic dynamics. Further 33 mutations
occurred in bit 5 until a mutation in 
bit 4 in time-step $t=10110$ generated a rapid movement towards a far
away region in the genotype space:
\begin{center}
\ttfamily
\begin{tabular}{ccc}
\vdots & \vdots & \vdots \\
10109 & 10001101000000010100 & 0.7060 \\
10110 & 100\underline{1}1101000000010100 & 0.7085 \\
10111 & 10011101000000010100 & 0.7081 \\
10112 & 10\underline{1}11101000000010100 & 0.7107 \\
10113 & 10111101000000010100 & 0.7098 \\
10114 & 1011110\underline{0}000000010100 & 0.7508 \\
\vdots & \vdots & \vdots \\
\end{tabular}
\end{center}
Further 40 mutations occurred between $t=10114$ and $t=11301$, until
finally, in $t=11302$, a new stable configuration was found:
\begin{center}
\ttfamily
\begin{tabular}{ccc}
\vdots & \vdots & \vdots \\
11301 & 11100010000000010000 & 0.7393 \\
11302 & 11100010000\underline{1}00010000 & 0.8014 \\
11303 & 11100010000100010000 & 0.8008 \\
11304 & 11100010000100010000 & 0.8001 \\
\vdots & \vdots & \vdots \\
\end{tabular}
\end{center}
This time, the dynamics had found a stable fix point. No further
mutations occurred until the simulation was aborted in time-step $t=40000$.

At this point, it is interesting to ask whether all limit cycles will 
be left for $t\rightarrow \infty$, or whether the
dynamics inevitably encounters a stable limit cycle eventually. A
possible answer to this question will be given in
Sec.~\ref{sec:ELnetworks}, where we present evidence for the existence
of a percolation transition in the region of intermediate $f_{\rm
  S}$. For $f_{\rm S}$ above the critical value, the dynamics would
sooner or later reach a stable fixed point, or a stable limit cycle,
whereas for $f_{\rm S}$ below the critical value, every limit cycle
which is part of the infinite (in the limit $N\rightarrow\infty$)
EL network will be eventually left.

\subsection{Short oscillation periods}

The oscillation period used in the previous section was relatively
long, if compared to the length $N$ of the bitstring. In such a situation,
a single site can be hit repeatedly by mutations during one
oscillation period. This generates complicated time correlations, which
are hard to tackle theoretically. The correlations can be neglected in
situations in which every site is, on average, hit at most once during
one oscillation period. This is assured if the number of time-steps
$T$ in one oscillation period is small compared to the number of bits $N$
in the string, 
\begin{equation}
  T \ll N\,.
\end{equation}
The case $f_{\rm S}=0$ can be treated analytically in this
regime. Again, we consider only a single frequency $\omega$, i.e.,
$\omega_i = \omega$ for all $i$. Let us begin with a bitstring of $N$
bits without epistatic interactions, i.e., with $K=0$. In this case,
the mutation of a single bit will change a single fitness contribution
$f_i(t)$ to a new form $f'_i(t)$. With what probability will the new
fitness contribution exceed the old one at time $t$? Since both
$f_i(t)$ and $f'_i(t)$ can be considered being chosen randomly from a
large set of possible functions, the probability equals the fraction
of the interval $[0;T)$ in which the function $f'_i(t)$ exceeds the
function $f_i(t)$. This fraction is easily calculated. We compute the
intersecting points $t_j^\ast$ from $f_i(t^\ast)=f'_i(t^\ast)$:
\begin{equation}
  \frac{1}{2}\left[\sin(\omega t^\ast + \delta_i) + 1 \right] = 
 \frac{1}{2}\left[\sin(\omega t^\ast + \delta'_i) + 1 \right]\,.
\end{equation}
With the use of a trigonometric identity, we find
\begin{equation}
  2\cos\left(\omega
  t^\ast+\frac{\delta_i+\delta'_i}{2}\right)
    \sin\left(\frac{\delta_i-\delta'_i}{2}\right)=0\,.
\end{equation}
Since $\delta_i=\delta'_i$ with probability 0, we assume that
$\delta_i$ and $\delta'_i$ are not equal, and find
\begin{eqnarray}
  t_1^\ast&=&\left(\frac{1}{4}-\frac{\delta_i+\delta'_i}{4\pi}\right)T\,,\\
 t_2^\ast&=&\left(\frac{3}{4}-\frac{\delta_i+\delta'_i}{4\pi}\right)T\,.
\end{eqnarray}
Hence, $f'_i(t)$ will exceed $f_i(t)$ half of the period, and
will be smaller than $f_i(t)$ during the other half. Therefore, the
probability $p_{\rm climb}$, with which a random mutation will lead to a
higher fitness, is $p_{\rm climb}=1/2$. The resulting bitstring's
motion in the genotype space is a random walk, in which jumps to neighboring
positions occur only with probability $1/2$.

The above result can easily be extended to higher $K$. The probability
$1/2$ actually stems from the symmetry that the fitness contributions
possess as trigonometric functions: 
\begin{equation}\label{eq:fitness_symm}
  f_i(t+T/2)-1/2=-\Big(f_i(t)-1/2\Big)\,.
\end{equation}
With Eq.~(\ref{eq:fitness_symm}), from $f'_i(t)>f_i(t)$ follows
$f'_i(t+T/2)<f_i(t+T/2)$. Half of the interval is dominated by $f_i(t)$,
the other half by $f'_i(t)$. A similar equation exists for the case
$K>0$. For positive $K$, a single mutation 
affects the fitness contributions of $K+1$ bits. $K+1$ fitness
contributions change. Without loss of generality, assume they are
numbered from 1 to $K+1$. Then, a mutation is accepted if the sum of
the changed fitness contributions exceeds the sum of the original
fitness contributions, i.e., if
\begin{equation}
  \frac{1}{K+1}\sum_{j=1}^{K+1} f'_j(t)\geq 
   \frac{1}{K+1}\sum_{j=1}^{K+1} f_j(t)\,.
\end{equation}
The symmetry property Eq.~(\ref{eq:fitness_symm}) of the single
fitness contributions can be transferred to the sum $\sum_{j=1}^{K+1}
f_j(t+T/2)/(K+1)$. With Eq.~(\ref{eq:fitness_symm}), we find
\begin{eqnarray}
   \frac{1}{K+1}\sum_{j=1}^{K+1} f_j(t+T/2) - 1/2
   &=&  \frac{1}{K+1}\sum_{j=1}^{K+1}\Big(-f_j(t)+1\Big)
             - 1/2\nonumber\\
  &=& -\left(\frac{1}{K+1}\sum_{j=1}^{K+1}f_j(t)  - 1/2\right)\,,
\end{eqnarray}
which is the exact equivalent of Eq.~(\ref{eq:fitness_symm}) for
positive $K$. Thus, for a given set of fitness contributions $f_j(t)$,
$f'_j(t)$, $j=1,2,\dots,K+1$, the sum of the contributions $f_j(t)$
will dominate half of the interval $[0;T)$, and it will be dominated
on the other half. Clearly, in the case $K>0$ the interval $[0;T)$ is divided
into more than 2 subintervals. However, the measure associated with the
subintervals in which $\sum f'_j(t)>\sum f_j(t)$ equals the measure
associated with the subintervals in which $\sum f'_j(t)<\sum f_j(t)$. 
The points for which 
both sums equal have measure zero, and can be neglected. The fact that
we now have to deal 
with several subintervals, instead of just 2, does not defeat the
argument given above for $K=0$. 
Therefore, the dynamics for $f_{\rm S}=0$ and $T\ll N$ is a random walk,
independent of the choice of $K$. For all $K$, the probability that
the next step in the random walk is taken is 1/2.

\begin{figure}
\begin{center}
\includegraphics{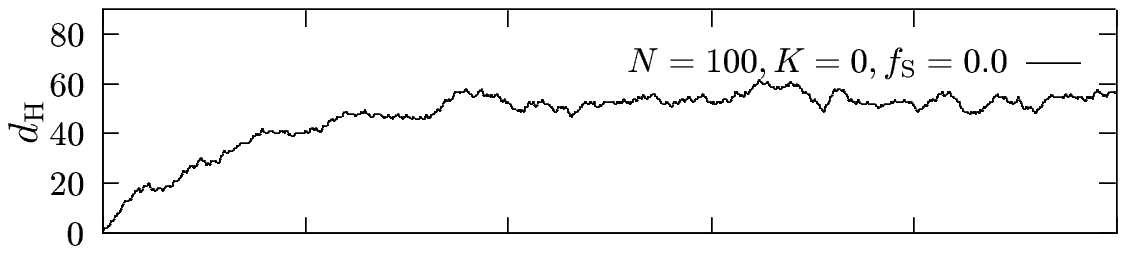}
\includegraphics{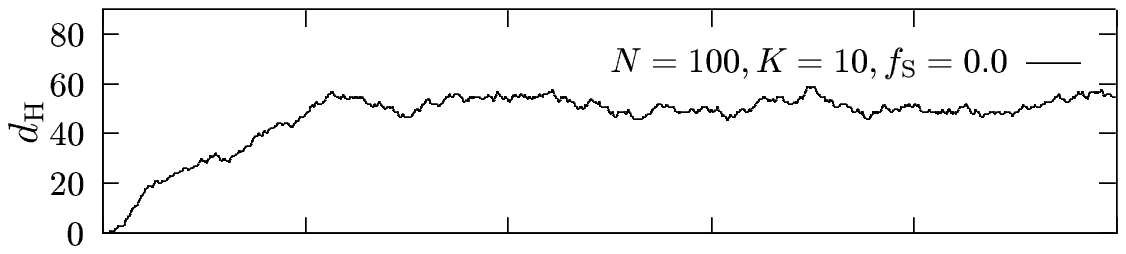}
\includegraphics{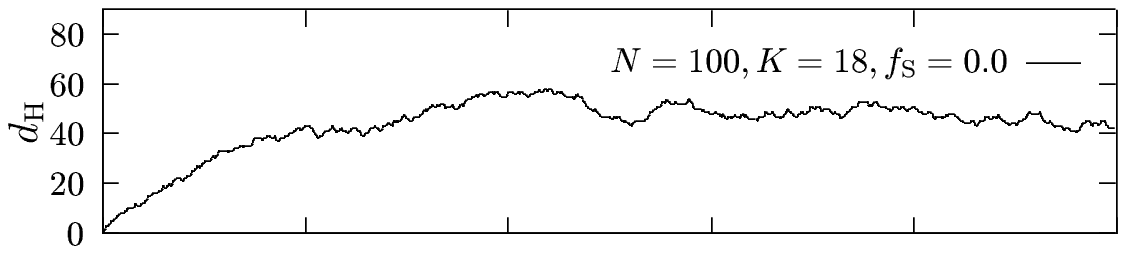}
\includegraphics{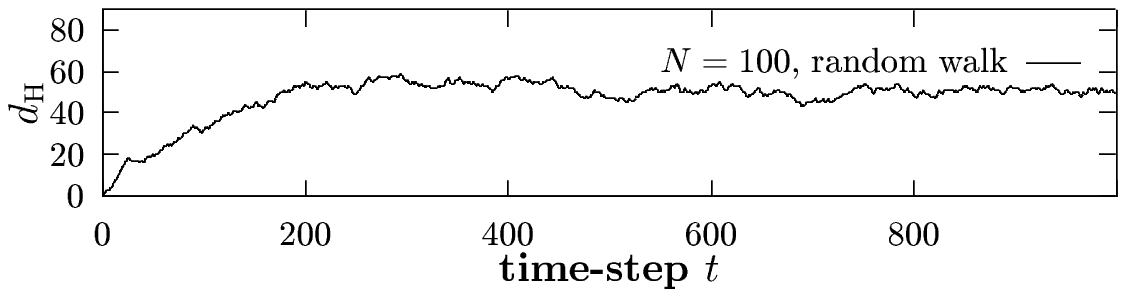}
\caption{The movement in Hamming distance $d_h$ away from the original
  string for 4 different experiments. The upper 3 graphs stem from
  oscillating $NK$ models. The oscillation period was $T=N=50$ in all
  cases. The bottom graph stems from a random walk with a probability
  of $1/2$ for resting one time-step.}
\label{fig:hamming}
\end{center}
\end{figure}

We have tested our analytical results in numerical experiments. The
adaptive walk behaves indeed very similar to a random walk for $f_{\rm
  S}=0$. In Fig.~\ref{fig:hamming}, we have
displayed the trajectories generated during the first 1000 time-steps
in 4 different computer experiments. Since it is hard to visualize a
high-dimensional trajectory, we plotted the Hamming distance $d_{\rm
  H}$ between the bitstring at time $t$ and the bitstring at time
0. Three of the experiments were done with oscillating $NK$-models,
the forth one was a simple random walk, used as a comparison. In all 3
experiments with oscillating landscapes, we used $T=N/2=50$. The
qualitative behavior is very similar in all 4 plots. In all cases, the
walker quickly moves away from the initial position in genome
space. For the rest of the run, the walker's Hamming distance from the
initial position fluctuates about $d_{\rm H}=50$. This is to be
expected for a random walk with $N=100$. On average, in two randomly
chosen sequences, half of the bits are the same. Therefore,
$d_{\rm H}$ fluctuates about $N/2$. 

\begin{figure}
\begin{center}
\includegraphics{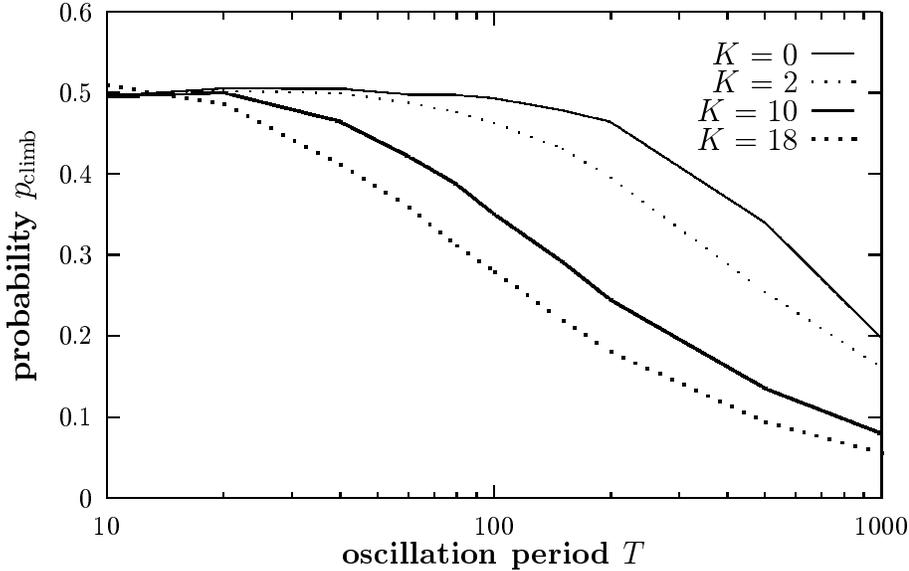}
\caption{The probability $p_{\rm climb}$ versus the oscillation period
  $T$ for different values of $K$. We used $N=100$ in all 4 cases.
  For small $T$, we observe a plateau  at $p_{\rm climb}=0.5$,
  whereas for large $T$, $p_{\rm climb}$ decays to zero. The decay
  happens the faster, the larger we set $K$.}
\label{fig:fS0Tvar}
\end{center}
\end{figure}

Some more insight can be gained from a direct measurement of $p_{\rm climb}$
for different period lengths $T$. The result of such a
measurement is presented in Fig~\ref{fig:fS0Tvar}.  For small $T$,
$p_{\rm climb}$ forms a plateau around 0.5, as it was predicted by our
analysis above. For large times $T$, the probability $p_{\rm climb}$
decays to zero.  

\subsection{EL networks and the percolation transition}
\label{sec:ELnetworks}

In this section, we will study the question of the existence of a
percolating regime for small $f_{\rm S}$. In the following, we will
only be interested in EL networks, i.e., we observe the dynamics only
at the beginning of each oscillation period, and disregard what is
going on in intermediate times. We will say an EL network percolates
if it consists of infinitely many points. This definition is similar
to the usual definition of the percolating cluster on the Bethe
lattice, and is the appropriate 
way to define percolation in high-dimensional
spaces~\cite{StaufferAharony92}. Clearly, it can be applied literally
only in the limit $N\rightarrow\infty$. However, the genotype space
grows so fast with increasing $N$ that this restriction can be
neglected. 

The study of EL networks in oscillating $NK$ landscapes is
computationally very demanding, since we have to go through the full
oscillation periods in the simulation. Hence, we restrict ourselves to
a single case we study as an example. We use $\omega_i=\omega=2\pi/T$
with $T=200$, as we have done in the previous sections. Moreover, we
consider only the case $N=20$.

\begin{figure}
\begin{center}
\setlength{\unitlength}{1cm}

\begin{picture}(14,8)
\put (0,4) {\includegraphics[width=6cm] {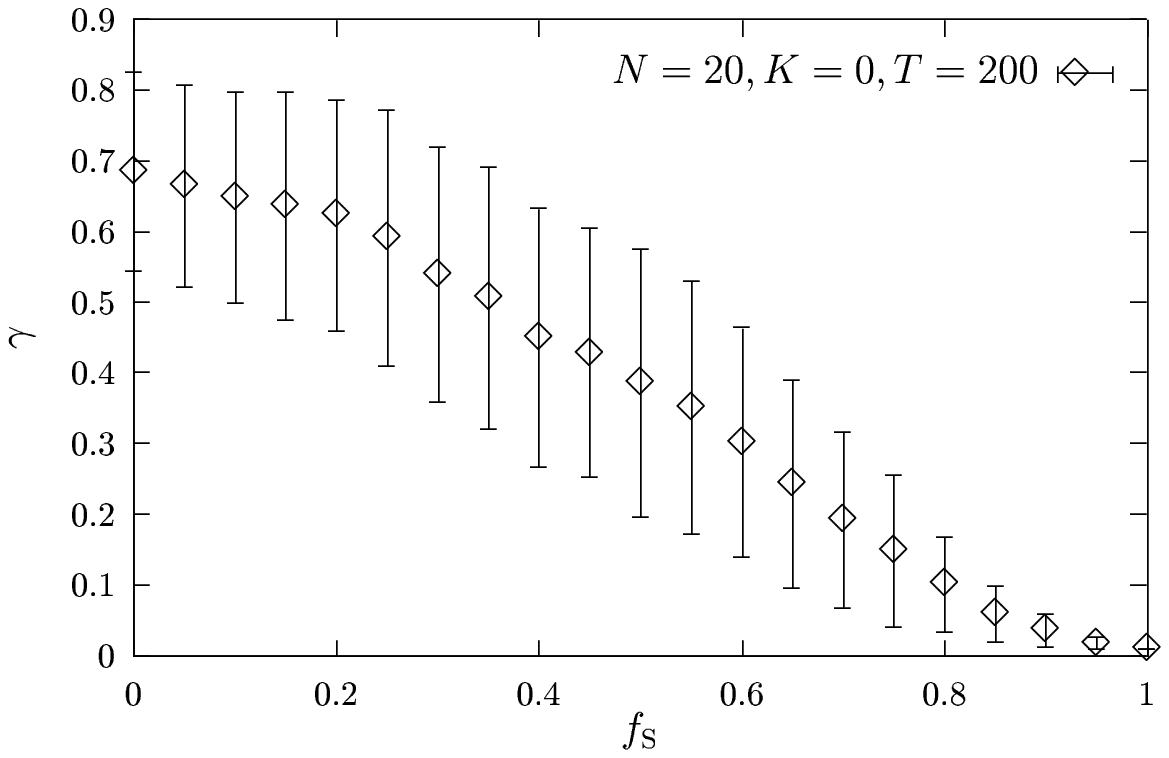}}
\put (7,4) {\includegraphics[width=6cm] {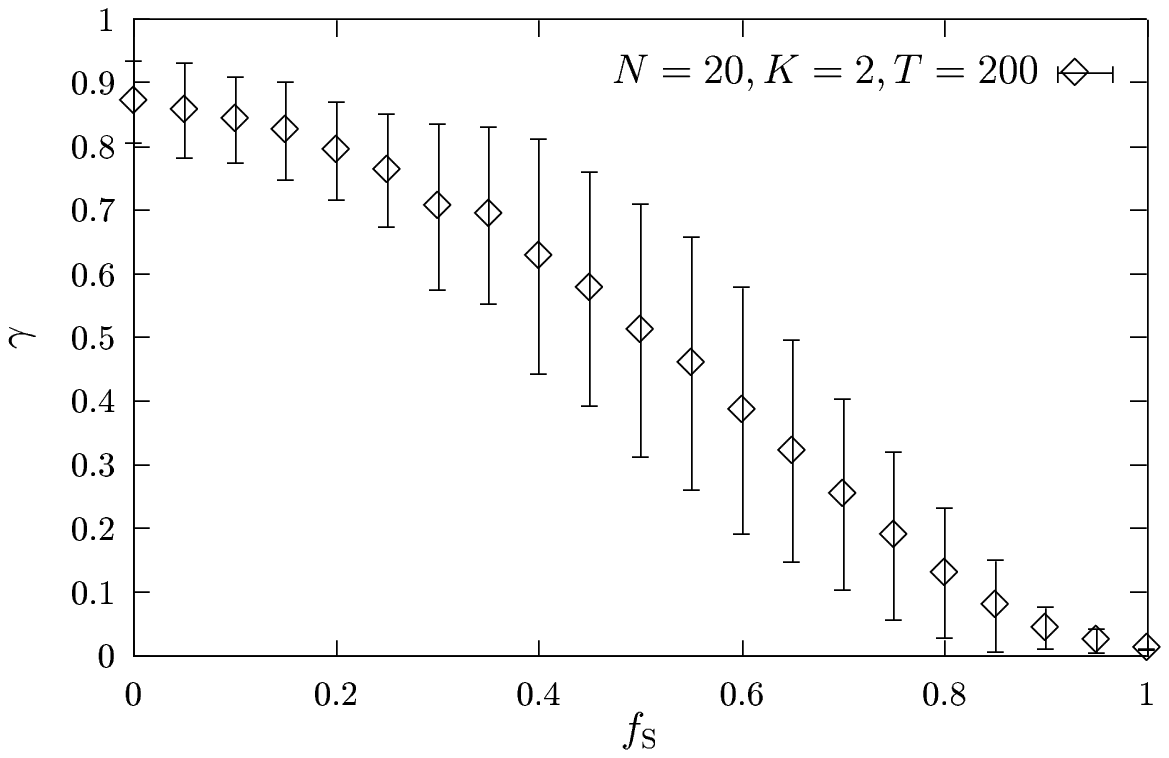}}
\put (0,0) {\includegraphics[width=6cm] {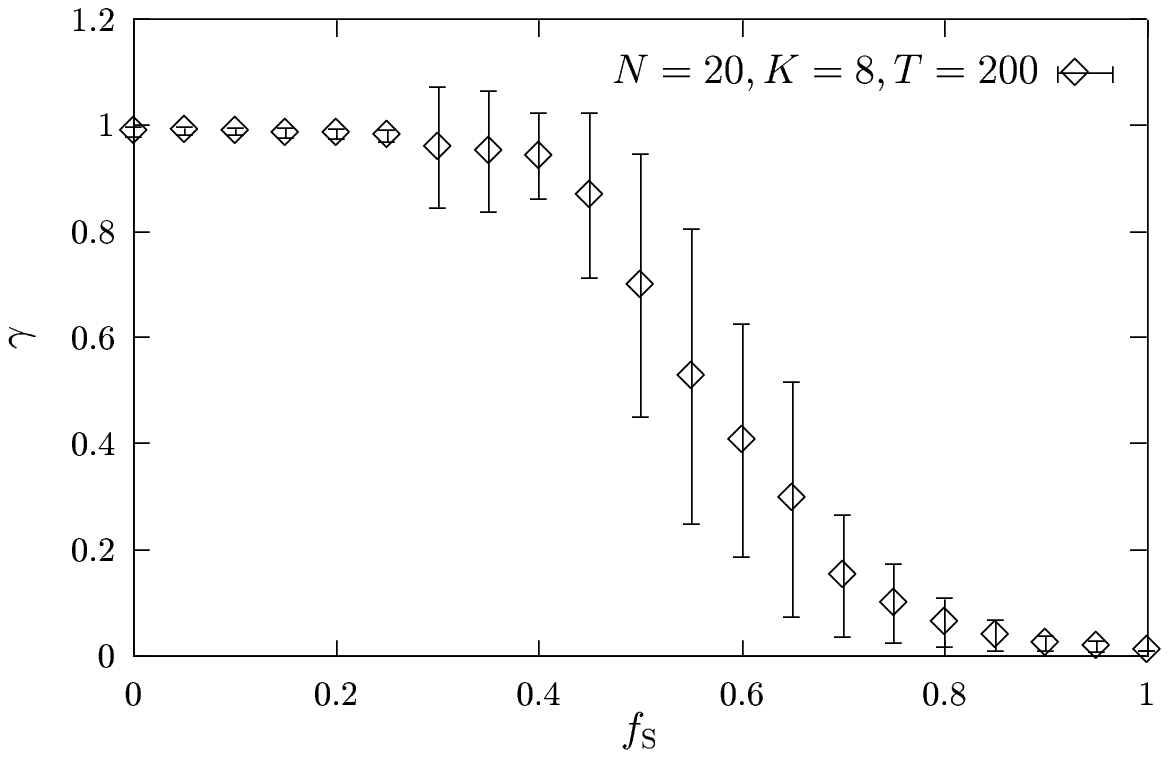}}
\put (7,0) {\includegraphics[width=6cm] {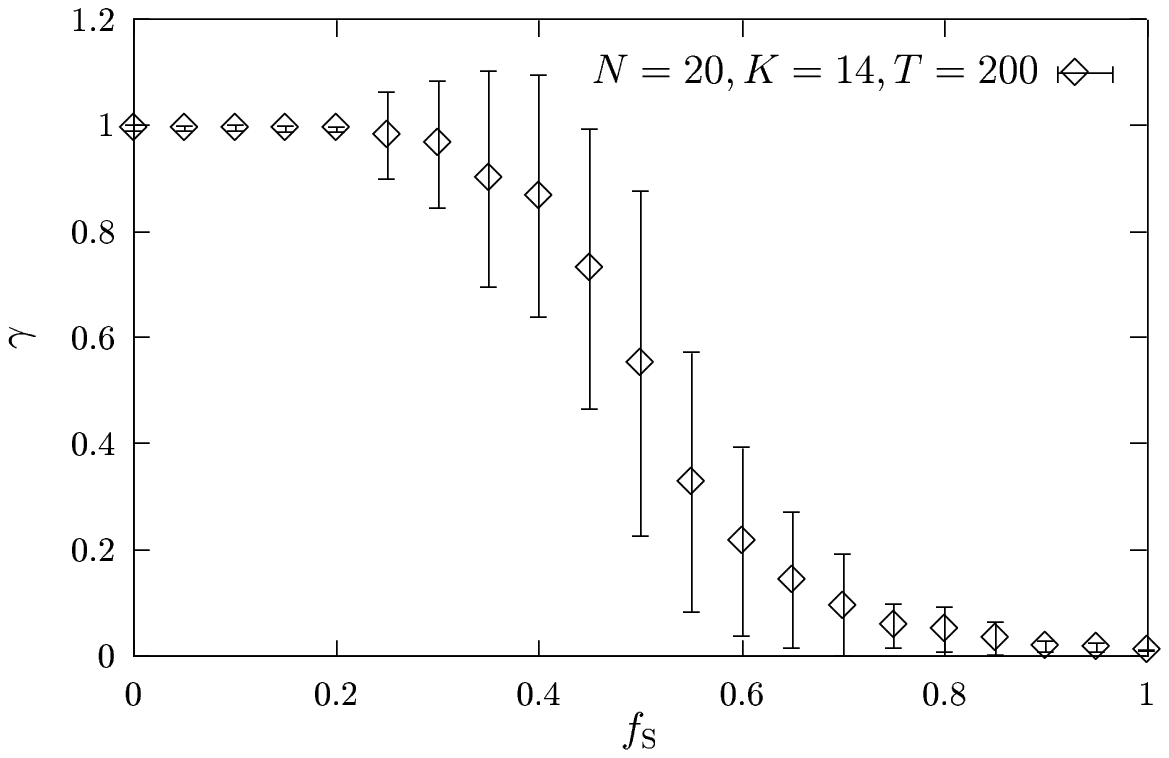}}
\end{picture}
\end{center}

\caption{The fraction of newly encountered genotypes at the beginning
  of each oscillation period in oscillating fitness landscapes. For small
  $f_{\rm S}$, a new genotype is encountered in almost every adaptive
  move. With increasing $f_{\rm S}$, the fraction of new genotypes
  decreases to zero. This happens the faster the larger the value of $K$.}
\label{fig:diversity}
\end{figure}

Figure~\ref{fig:diversity} shows the fraction $\gamma$ of new genotypes
among all the genotypes encountered at the beginning of each
oscillation period. This is a measure for the size of an EL network. A
value near 1 means a new genotype has been encountered in almost every
oscillation period. On the other hand, a value near 0 means the
network's size is small, thus confining the adaptive walk in a limited
region of the genotype space. In the limit of infinitely many
oscillation periods, only percolating networks can have a positive
$\gamma$, whereas finite networks yield $\gamma=0$. Therefore,
$\gamma$ is a proper order parameter to indicate a percolation
transition. Clearly, in numerical experiments the number of
oscillation periods over which the measurement is taken is finite, and
therefore we will observe a positive $\gamma$ even in the
non-percolating regime. In the case of Fig.~\ref{fig:diversity}, for
example, the value $\gamma$ was obtained from averaging over 60
adaptive walks, each on a different fitness landscape. Every adaptive
walk endured 200 oscillation periods. The error bars present the
standard deviations of the single measurements.

Let us begin the discussion of Fig.~\ref{fig:diversity} with the two
graphs in the bottom row, for $K=8$ and $K=14$. In both cases,
$\gamma=1$ for small $f_{\rm S}$, and $\gamma=0$ for $f_{\rm S}\approx
1$. The standard deviations are very small in both limiting
regimes. In the region about $f_{\rm S}\approx 0.5$, a sharp drop in
$\gamma$ can be observed, accompanied with an enormous increase in the
error bars. The large error bars are a sign for critical fluctuations,
which are observed in 2nd order phase transitions. The picture
is not as clear in the upper row of Fig.~\ref{fig:diversity}, for
$K=0$ and $K=2$. The error bars are large for the whole range of
$f_{\rm S}$, and $\gamma$ does not reach the value 1 for $f_{\rm
  S}=0$. This implies a percolating El network, if it exists for small
$K$ (below we will give an argument why this is probably not the case
for $K=0$), does not span the whole genotype space in this case, even
in the limit $f_{\rm S}=0$. The transition from the percolating to
the non-percolating regime has to be much softer, and the critical
value $f_{\rm S}^\ast$ cannot be estimated from our measurements. 

Measurements with an extremely increased accuracy are necessary to
determine the critical point $f_{\rm S}^\ast$ exactly, and to decide
whether a percolation transition does exist at all for small $K$. This
work is currently being undertaken, and will be presented elsewhere.

The measurements in Fig.~\ref{fig:diversity} reveal another surprising
fact. For small $f_{\rm S}$, $\gamma$ is increased with increased
ruggedness of the landscape. The ``Fujiyama'' landscape $K=0$ has a
much smaller $\gamma$ than the relatively rugged landscape with
$K=8$. This means ruggedness promotes the movement in the genotype
space for the oscillating regime, a situation completely opposite to
the case of fixed fitness landscapes, where ruggedness is regarded as
an impediment to the movement in the genotype space. We can understand
this observation in the ``adiabatic'' limit, where changes in the
fitness landscape happen so slowly that the adaptive walk can always
reach the next local optimum before another change happens. In this
limit, the EL network of the Fujiyama landscape degenerates to a
single point. On the other hand, the completely random Derrida
landscape~\cite{Derrida81} which we obtain for $K=N-1$ presents a
multitude of local optima, and the changes 
in the fitness landscape provide the opportunity to hop from one local
optimum to another in a random fashion during the oscillation periods.

\section{Conclusions}

Evolution in a slowly changing environment follows a dynamics very
different from the situation in a fixed environment. Environmentally
guided drift drives genes out of local optima, and drags them around
in the genotype space. The evolution generally proceeds on $\epsilon$-EL
networks. However, transitions from one $\epsilon$-EL network to another
occur sporadically. In this work, we have presented
evidence for the existence of a giant $\epsilon$-EL network in the
limit of a vanishing amount of static fitness
contributions. Consequently, in this limit the whole genotype space
can be transversed by means of environmentally guided drift.
 Hence, the guided drift can provide -- in absence of any
neutral pathways in the fitness landscape -- an efficient mechanism to
generate constantly new genotypes, albeit at every single
point in time the system seems to be trapped in a local optimum.
In Sec.~\ref{sec:ELnetworks}, we could show that the efficiency of the
environmentally guided drift is related to the ruggedness of the
landscape. A more rugged landscape provides more opportunities to move
around under environmental changes than a landscape with only a
few peaks. Consequently, the rugged landscapes observed in protein
evolution~\cite{BennerEllington88} could accelerate protein evolution
in an ever changing environment, instead of hindering it. 

Above the percolation transition, the $\epsilon$-EL networks are
relatively small. The dynamics in this regime is dominated by noisy
limit cycles. The
system goes through several noisy limit cycles until a stable
limit cycle, or a stable fixed point, is reached.

The model studied in the present paper, i.e., the family of
oscillating $NK$-landscapes, is certainly too simplistic
to be accounted for as a realistic model of the
evolution of a species in a changing environment. In particular, it
can be argued whether oscillating fitness contributions, taking on
all values between zero and one during one oscillation period, are
justified. Nevertheless, such a simple model can be used to get a
first impression of how evolution in a slowly changing environment
could behave. Furthermore, a simple model can guide the development of
tools and concepts that can then be
applied to more complicated situations. The idea of
environmentally linked networks is such a concept. EL networks can be
defined on any fitness landscape with recurrent states. The 
restriction to such fitness landscapes may seem narrowing
on the first sight, but it is justified in a number of reasonable
situations, like the evolution of microorganisms under the change of
seasons, as mentioned in Sec.~\ref{sec:env_guided_drift}. Generally,
two states of a system at different points in
time can only be compared under similar boundary
conditions. Therefore, a condition like 
Eq.~(\ref{eq:FL_equal_states}) is necessary to draw meaningful
comparisons. In particular, in evolution two different adaptive
solutions can only be compared if the problem to solve is identical.

In future work, it should be interesting to study the percolation
transition in more detail, and to determine for what $K$ a percolating
regime actually exists. Furthermore, higher oscillation modes
should be considered and the influence of the string length $N$ and
the period length $T$ should be investigated. 
Investigations similar to the ones we have presented in this work can also
be done for other time-periodic fitness landscapes. 

\section {Acknowledgments}
I would like to thank Steven Benner, who
made me aware of the strong constraints on the evolution of
proteins imposed  by natural selection, and of the environmental
changes' consequent importance to innovation in evolution.

\bibliographystyle{plain}
\bibliography{/home/wilke/tex/bibtex/bibdat,/home/wilke/tex/bibtex/mypub}

\end{document}